\documentclass[aps,pre,twocolumn]{revtex4}
\usepackage{graphics}
\usepackage{graphicx}

\begin{document}


\title{L\'evy flights on a comb and the plasma staircase}


\author{Alexander~V.~Milovanov${}^{1,2}$ and Jens~Juul~Rasmussen${}^{3}$}

\affiliation{${}^1$ENEA National Laboratory, Centro~Ricerche~Frascati, I-00044 Frascati, Rome, Italy}
\affiliation{${}^2$Space Research Institute, Russian Academy of Sciences, 117997 Moscow, Russia}
\affiliation{${}^3$Physics Department, Technical University of Denmark, DK-2800 Kgs.~Lyngby, Denmark}




\begin{abstract} We formulate the problem of confined L\'evy flight on a comb. The comb represents a sawtooth-like potential field $V(x)$, with the asymmetric teeth favoring net transport in a preferred direction. The shape effect is modeled as a power-law dependence $V(x) \propto |\Delta x|^n$ within the sawtooth period, followed by an abrupt drop-off to zero, after which the initial power-law dependence is reset. It is found that the L\'evy flights will be confined in the sense of generalized central limit theorem if (i) the spacing between the teeth is sufficiently broad, and (ii) $n > 4-\mu$, where $\mu$ is the fractal dimension of the flights. In particular, for the Cauchy flights ($\mu = 1$), $n>3$. The study is motivated by recent observations of localization-delocalization of transport avalanches in banded flows in the Tore Supra tokamak and is intended to devise a theory basis to explain the observed phenomenology.
\end{abstract}



\maketitle

\section{Introduction} 
In recent investigations of zonal flow phenomena in magnetized plasma by means of high-resolution ultrafast-sweeping X-mode reflectometry in the Tore Supra tokamak, spontaneous flow patterning into a quasi-regular sequence of strong and lasting jets interspersed with broader regions of turbulent (typically, avalanching) transport has been observed \cite{DP2015,Hornung,DP2017}. The phenomenon was dubbed ``plasma staircase" by analogy with its notorious planetary analogue \cite{McIn}. The plasma staircase has been referred as an important self-organization phenomenon of the out-of-equilibrium plasma, which had pronounced effect on radial transport and the quality of confinement. Detailed analyses  (both experimental and numerical based on gyrokinetic calculations) have identified the plasma staircase as a weakly collisional, meso-scale \cite{Meso} dynamical structure near the state of marginal stability of the low confinement mode plasma \cite{Hornung,DP2017}. 

The comprehension of the plasma staircase \cite{DP2015} has both fundamental and practical significance. From a scientific perspective, the plasma staircase represents a fascinating dynamical system in which kinetic and fluid nonlinearities may operate on an equal footing. In the practical perspective, the plasma staircase raises the important problem of avalanche-zonal flow interaction \cite{Hornung,DP2017}, which may be key to control the dynamic confinement conditions in magnetic fusion devices, tokamaks and stellarators. On top of this, the fact that a significant portion, if not a vast majority, of avalanches have been confined within the staircase steps \cite{DP2017} is by itself a challenge, since the plasma avalanches being {spatially extended} transport phenomena behave dynamically nonlocally, and their ``localization" within a transport barrier is not at all obvious. Mathematically, this revives the long-standing problem of the {\it confined L\'evy flight}, which has attracted attention in the literature previously (e.g., Refs. \cite{Chechkin2002,Ch2003,Ch2004,Klafter2004,Ch2005,Ch2007}).     

In this paper, we adapt the general problem of confined L\'evy flight \cite{Ch2003,Klafter2004} for staircase physics and show that the transport avalanches may be localized, if (i) the staircase jets are spatially separated, as they prove to be \cite{Hornung,DP2017}, and (ii) at each step of the staircase the gradients are sharp enough in that the potential function grows faster with distance than a certain critical dependence (cubic when modeled by a power-law). If the growth is slower than this, then the avalanches are {\it not} localized in that there is an important probability of finding the L\'evy flyer outside the transport barrier. More so, we find that in the confinement domain there may occur at least three different types of avalanches, which we call, respectively, {\it white } swans, {\it black} swans \cite{Taleb} and {\it dragon kings} \cite{Sornette}, and that the white swans may ``mutate" into the black swan species past the intermediate {\it grey-swan} family found at the point of cubic dependence. This gives rise to some features of bifurcation, which might be identifiable in the experiment. This observation opens a new perspective on ``smart" plasma diagnostics in tokamaks using plasma self-organization \cite{DP2015,Hornung,DP2017}.  

The paper is organized as follows. In Sec. II, we introduce an idealized transport model, which we arguably name {\it L\'evy flights on a comb}, and which is motivated by the challenges discussed above. The model, which is derived in Sec. II-B using the idea of transition probability in reciprocal space \cite{PLA}, is intended to mirror the observed behaviors \cite{DP2015,Hornung,DP2017} and, most importantly, provide a practical criterion for the phenomena of localization-delocalization of avalanches in the presence of zonal flows. We discuss the various aspects of this model in Secs. III and IV, which focus on, respectively, space scale separation issues and the size distribution of avalanches. The latter is shown to be inverse power-law for both the white and black swans, but with different drop-off exponents, making it possible to differentiate between the species. We conclude the paper in Sec. V with a few remarks.   

\section{The model}
We represent the plasma staircase as a periodic lattice, a comb, looking along the coordinate $x$; the latter represents the radial direction in a tokamak. The comb, with its sharp teeth, mimics the very concentrated jets in the cross-section of poloidal flows, which define both the periodic structure and the spatial step of the staircase (see Fig.~1). $j$ is a natural number and counts the teeth of the comb along the $x$ axis starting from the inner ones, such that $x_j$ would be the location of the $j$-th tooth in radial direction. The spacing between the neighboring teeth is $\Lambda = |x_{j+1} - x_j|$ and is assumed to not depend on $j$. Also we assume that the number of teeth is statistically large (i.e., $j_{\max} \gg 1$), and that $\Lambda$ is much smaller than the tokamak minor radius. For each pair of neighboring teeth, with the radial locations at $x_j$ and $x_{j+1}$, we introduce a potential function, $V(\Delta x)$, which grows with the departure $\Delta x$ from $x_j$ as a power-law, i.e., $V(\Delta x) \propto |\Delta x|^n$. The exponent $n$ is not necessarily integer. We assume that $n$ is larger than 2, so that $V(x)$ is concave, with the vanishing first and second derivatives for $x\rightarrow +0$. This condition is needed for ``stability" of the ensuing power-law like probability distributions and will be illustrated below. For $x_{j} < x < x_{j+1}$, $V(x)$ is continuous, with the boundary condition $V(x_{j} +0) = 0$. When $x$ approaches $x_{j+1}$ from the left, the function $V(x)$ reaches its maximal allowed value $V_{\max} = V(\Lambda)$ at $x = x_{j+1} -0$. Crossing the tooth at $x = x_{j+1}$, its value is disconnected, and is turned down to zero at $x=x_{j+1} +0$. Then the power-law dependence $\propto |\Delta x|^n$ is reset for $x > x_{j+1}$ until the next tooth is met, etc. The abrupt drop-off to zero in the $V(x)$ dependence at $x=x_{j+1} +0$ implies there is a strong repulsive force acting on a passive particle at the right border of each tooth. This favors transport towards the ever increasing values of $x$ (i.e., towards larger radial locations in the direction of the scrape-off layer in a tokamak). In real magnetic confinement systems, this behavior involves the shape of the background density and temperature profiles, as well as the relevant toroidicity effects \cite{Horton}. The potential function $V(x)$ represents the barriers to radial transport. Such barriers occur spontaneously via self-organization of the tokamak plasma under certain conditions \cite{Diamond}. The form and characteristics of the $V(x)$ dependence are rooted in the basic physics of vortical flows and the notion of potential vorticity \cite{McIn,Madsen}. Note that the teeth of the $V(x)$ function are, by their construction, strongly shaped and not symmetric; when drawn to a graph, the periodic dependence in $V(x)$ looks like a saw. The ``abrupt" drop-offs to zero at $x=x_{j+1} +0$ should be taken with a grain of salt, and it is understood that there is a finite spatial spread there, which is determined by finite plasma viscosity. 

\begin{figure}
\includegraphics[width=0.55\textwidth]{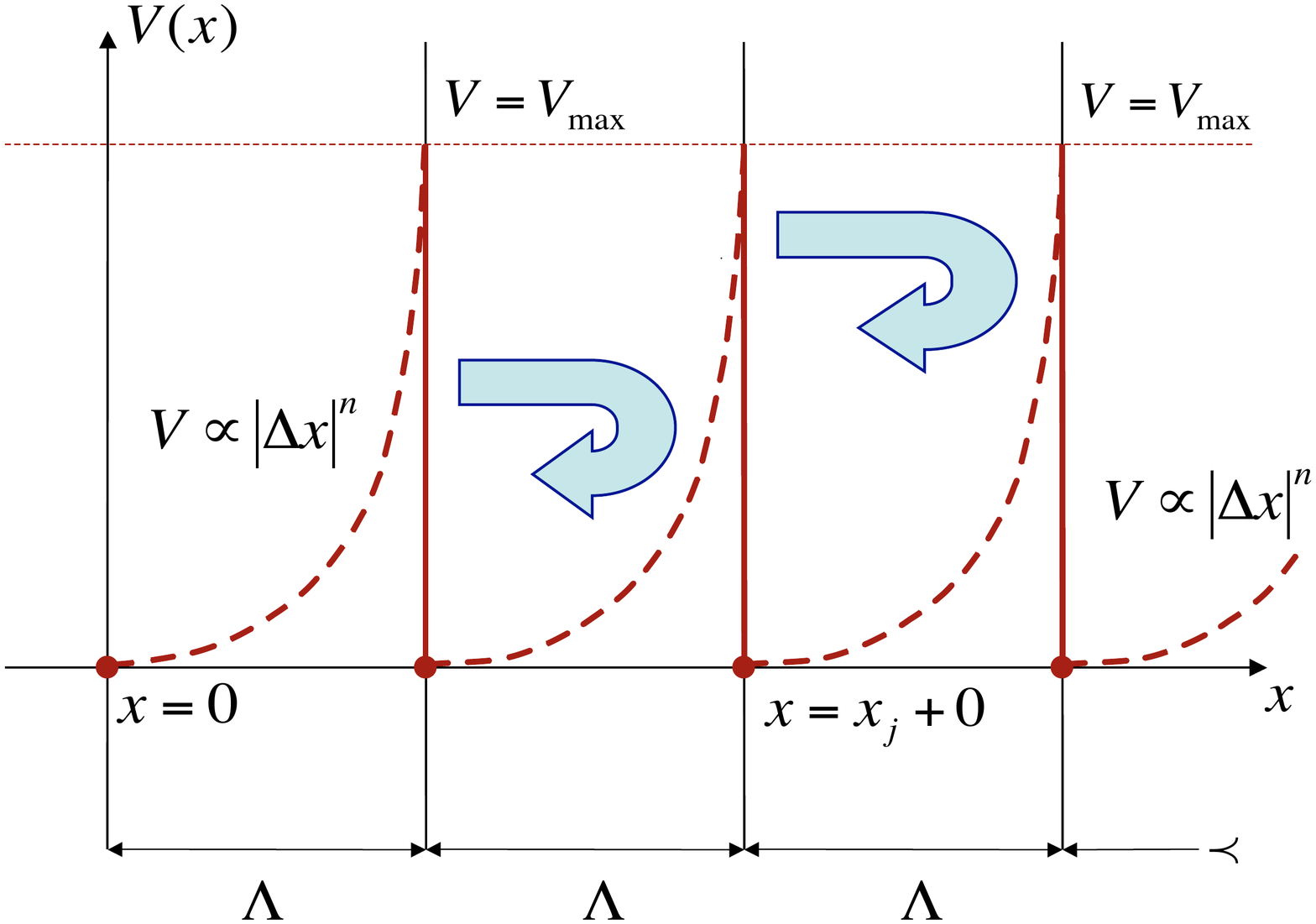}
\caption{\label{} The comb model. The staircase jets go perpendicular to the figure plane and are marked by fat dots at $x=x_j$. The sawtooth effect is modeled by the power-law dependence $V(\Delta x) \propto |\Delta x|^n$ at each step of the staircase. We are interested in finding the conditions permitting to localize the avalanches (U-turn arrows) in-between the staircase steps.}
\end{figure}

The question we pose now is whether an avalanche, emitted at the radial location $x=x_j +0$, can be confined by a potential field $V(x)\propto (x-x_j)^n$ for $x \rightarrow +\infty$ (and what would {\it confined} mean in that case). This setting assumes that the spacing between consecutive teeth of the comb is very broad, permitting to neglect possible interferences between the various pieces of the saw-like $V(x)$. With these implications in mind, we just single out one piece by allowing $\Lambda\rightarrow+\infty$. This idealization does not influence the final conclusions concerning localization-delocalization of avalanches, but appreciably simplifies the analysis. Without loss in generality, we also set $x_j = 0$, and we omit the index $j$ hereafter to simplify notations. To this end, $V(x) \propto x^n$ for $x > 0$, with $n > 2$ (by far $n$ remains a free parameter of the model and will be conditioned later). Dynamically, the assumption that the spacing $\Lambda$ is large, i.e., $\Lambda\rightarrow+\infty$, is equivalent to requiring that the kinetic energy, involved in an avalanche event, is much smaller than $V_{\max} = V(\Lambda)$. In a self-regulating nonlinear system, that would be reasonably well satisfied, since the avalanches, absorbed by the transport barriers, deliver momentum to the poloidal flows (via the turbulent Reynolds stress), which in turn enhances the strength of the barrier \cite{Xu}. Further concerning the $\Lambda$ value, we estimate this as the Rhines length in the coupled avalanche-zonal flow system. In fluid dynamics, the Rhines length \cite{Rhines} determines the upper bound on the size of vortical structures in the flow. In drift-wave turbulence, the analogue Rhines length is introduced \cite{Naulin}, which is shown to scale with the $E\times B$ velocity (as a square-root of this). In this regard, $\Lambda$ is the level of electrostatic drift-wave turbulence driving the staircase, so that $\Lambda\rightarrow+\infty$ would imply that the turbulence intensity is actually very high.  

\subsection{Basic equations, nonlocality, and the Cauchy limit}
As a general approach, we consider a transport model of the Fokker-Planck type, with due modifications accounting for the presence of transport avalanches, on the one hand, and the effect of external potential field, $V(x)$, on the other hand. The model, which has been devised for magnetically confined plasma in Ref. \cite{PLA}, may be summarized in terms of the following kinetic equation for the probability density $f = f (x, t)$ to find a passive tracer at time $t$ at point $x$:
\begin{equation}
\left[\frac{\partial}{\partial t} - \frac{1}{\eta} \frac{\partial}{\partial x} V^{\prime}(x) \right] f (x, t) = \hat{T} f (x, t) + \hat{S}_{\pm}[f (x, t)], 
\label{MoDeL+} 
\end{equation} 
where $V^{\prime}(x) = d V(x)/dx$ is the gradient of the sawtooth field along the $x$ axis; $-V^{\prime}(x)$ is the radial force felt by the particle and is responsible for the convection term in Eq.~(\ref{MoDeL+}); $\eta$ is viscosity (in the fluid sense) and determines the actual finite spread in the $V(x)$ jumps (neglected in the idealized model); $\hat{S}_{\pm}[f (x, t)]$ is the source/sink term, which is defined as a functional on $f (x, t)$; and 
\begin{equation}
\hat{T} f (x, t) = D \frac{\partial^2}{\partial x^2} f (x, t) + \frac{\partial^2}{\partial x^2} \Psi_\mu (x, t)
\label{DL} 
\end{equation} 
is a combination of Gaussian diffusion (the first term on the right-hand-side, identified by the coefficient $D$) and nonlocal diffusion accounting for the avalanche processes in the medium (this term is marked by the index $\mu$ and is identified by the nonlocal function $\Psi_\mu (x, t)$ to be quantified below). The combined avalanche-diffusion model in Eqs.~(\ref{MoDeL+}) and~(\ref{DL}) is {derived} below based on a Markov evolution equation for the probability density $f (x, t)$, using random walks and the notion of transition probability in Fourier space (see Sec. II-B). The assumption of Markovianity says we shall neglect any possible trapping phenomena at the staircase steps. The Gaussian term in Eq.~(\ref{DL}) stands for the familiar collisional diffusion in a weakly collisional plasma. This term may naturally be extended, so that it also includes the quasilinear (collisionless) diffusion by wave-particle interactions \cite{Sagdeev}. The nonlocal term in Eq.~(\ref{DL}) accounts for the presence of the coherent structures in the medium, that is, the avalanches. It is understood that the avalanches propagate radially on a very fast time scale (much faster than the corresponding diffusive times) and are characterized by a velocity close to the ion acoustic speed \cite{DP2017}. An account on the observation and quantitative characterization of avalanche events in a magnetically confined plasma can be found in Ref. \cite{Politzer}. As the avalanches can trap and convect particles, they may cause their sudden displacements in radial direction occurring at about the sonic speeds. Such processes would be virtually instantaneous when compared to the microscopic diffusion processes (collisional or quasilinear). We consider these sudden radial displacements caused by the avalanches as the Cauchy flights along the $x$ axis. The Cauchy flights are partial case of more general L\'evy flights and correspond to the limit $\mu\rightarrow 1$ in the L\'evy fractional diffusion equation (e.g., Refs. \cite{Ch2007,Klafter2004,Klafter}) 
\begin{equation}
\frac{\partial}{\partial t} f (x, t)  = K_\mu \frac{\partial^2}{\partial x^2} \frac{1}{\Gamma_\mu} \int_{-\infty}^{+\infty}\frac{f (x^\prime, t)}{|x-x^\prime|^{\mu - 1}} dx^\prime.
\label{FDE} 
\end{equation} 
The integro-differential operator on the right-hand-side of Eq.~(\ref{FDE}) is known as the Riesz fractional derivative and incorporates the nonlocal features of L\'evy flights \cite{Ch2007,Klafter} via a convolution with a power-law. Also in Eq.~(\ref{FDE}) one encounters $K_\mu$, the transport coefficient, which carries the dimension cm$^\mu\,\cdot\,$sec$^{-1}$; as well as the normalization parameter $\Gamma_\mu = - 2\cos(\pi\mu/2)\Gamma (2-\mu)$, which occurs by splitting the improper integration in Eq.~(\ref{FDE}) into two Riemann-Liouville integrals, i.e., $\int_{-\infty}^{+\infty} = \int_{-\infty}^{x} + \int_{x}^{+\infty}$. Further, $\mu$ is the {\it fractal dimension} of L\'evy flights \cite{Klafter}. This is a numerical parameter lying between the two integer limits, i.e., $\mu = 1$ (posed by topological connectedness of the L\'evy flight trajectories) and $\mu = 2$, for which the nonlocal properties vanish. Note, in this regard, that the normalization parameter $\Gamma_\mu \rightarrow +\infty$ for $\mu\rightarrow 2$ (due to the divergence of the gamma function), saying it is solely the Gaussian diffusion term in Eq.~(\ref{DL}) that survives in this limit. For $\mu = 1$, the integro-differentiation on the right-hand-side of Eq.~(\ref{FDE}) reduces (via the degeneration of the normalization parameter) to the Hilbert transform operator \cite{Mainardi}, leading to the following simplified kinetic equation for Cauchy flights in an infinite space
\begin{equation}
\frac{\partial}{\partial t} f (x, t)  = - K_1 \frac{1}{\pi} \frac{\partial}{\partial x} \int_{-\infty}^{+\infty}\frac{f (x^\prime, t)}{x-x^\prime} dx^\prime,
\label{FDE+} 
\end{equation} 
where $K_1 = \lim_{\mu\rightarrow 1} K_\mu$. Dynamically, the limit $\mu\rightarrow 1$ serves to emphasize that the Cauchy flights are kind of very fast, ballistic displacements along the $x$ axis, and as such they mirror the observed avalanche phenomenology at the staircase steps \cite{Hornung,DP2017}. For $1\leq \mu < 2$, the algebraic kernel in Eq.~(\ref{FDE}) characterizes the nonlocal nature of transport avalanches. Note that the Fickian transport paradigm that fluxes are decided by local gradients \cite{Fick} does not apply here. The fact that the nonlocal properties are inherently present in the coupled avalanche-zonal flow system has been demonstrated in Ref. \cite{Meso} based on flux-driven gyrokinetic \cite{Hahm} computations, using generalized heat transfer integrals and the heuristic idea of ``influence length." A clear evidence of nonlocal effects in tokamak plasma was provided by perturbative experiments \cite{Mantica,Mantica_etal} with plasma edge cooling and heating power modulation, indicating anomalously fast transport of edge cold pulses to plasma core, not compatible with major diffusive time scales \cite{PLA,Pulse,Hariri}. Recent progresses on experimental analysis and theoretical models for nonlocal transport (non-Fickian fluxes in real space) are reviewed in Ref. \cite{Itoh}.  

\subsection{Derivation of the nonlocal term}
Before we proceed with the main topics of this study, we wish to illustrate the derivation of the transport model in Eqs.~(\ref{MoDeL+}) and~(\ref{DL}) above, using the idea of transition probability in reciprocal space (Ref. \cite{PLA}; references therein). For this, consider a Markov (memoryless) stochastic process defined by the evolution equation
\begin{equation}
f(x, t+\Delta t) = \int_{-\infty}^{+\infty} f(x-\Delta x, t) \psi (x, \Delta x, \Delta t)d\Delta x,
\label{1} 
\end{equation}
where $f (x, t)$ is the probability density of finding a particle (random walker) at time $t$ at point $x$, and $\psi (x, \Delta x, \Delta t)$ is the transition probability density of the process. Note that the ``density" $\psi (x, \Delta x, \Delta t)$ is defined with respect to the increment space characterized by the variable $\Delta x$. It may include a parametric dependence on $x$, when non-homogeneous systems are considered. Here, for the sake of simplicity, we restrict ourselves to the homogeneous case, and we omit the $x$ dependence in $\psi (x, \Delta x, \Delta t)$ to enjoy
\begin{equation}
f(x, t+\Delta t) = \int_{-\infty}^{+\infty} f(x-\Delta x, t) \psi (\Delta x, \Delta t)d\Delta x.
\label{2} 
\end{equation} 
Then $\psi (\Delta x, \Delta t)$ defines the probability density of changing the spatial coordinate $x$ by a value $\Delta x$ within a time interval $\Delta t$ independently of the running $x$ value. The integral on the right of Eq.~(\ref{2}) is of the convolution type. In the Fourier space this becomes
\begin{equation}
\hat f(k, t+\Delta t) = \hat f(k, t) \hat \psi (k, \Delta t),
\label{3} 
\end{equation} 
where the integral representation  
\begin{equation}
\hat \psi (k, \Delta t) = \hat \mathcal{F} \{\psi (\Delta x, \Delta t)\} \equiv \int_{-\infty}^{+\infty} \psi (\Delta x, \Delta t) e^{ik\Delta x} d\Delta x
\label{Fourier} 
\end{equation} 
has been used for $\hat \psi (k, \Delta t)$, and similarly for $\hat f(k, t)$. Letting $k\rightarrow 0$, it is found that 
\begin{equation}
\lim_{k\rightarrow 0}\hat \psi (k, \Delta t) = \int_{-\infty}^{+\infty} \psi (\Delta x, \Delta t) d\Delta x.
\label{F2} 
\end{equation} 
The improper integral on the right hand side is nothing else than the probability for the space variable $x$ to acquire {\it any} increment $\Delta x$ during time $\Delta t$. For memoryless stochastic processes without trapping, this probability is immediately seen to be equal to 1, that is, the diffusing particle takes a displacement anyway in any direction along the $x$-axis. Therefore,
\begin{equation}
\lim_{k\rightarrow 0}\hat \psi (k, \Delta t) = 1.
\label{F2+} 
\end{equation} 
We consider $\hat \psi (k, \Delta t)$ as the average time-scale- and wave-vector-dependent transition ``probability" or the characteristic function of the stochastic process in Eq.~(\ref{2}). In general, $\hat \psi (k, \Delta t)$ can be due to many co-existing, independent dynamical processes, each characterized by its own, ``partial" transition probability, $\hat \psi_j (k, \Delta t)$, $j=1,\dots n$, making it possible to expand
\begin{equation}
\hat \psi (k, \Delta t) = \prod_{j=1}^n \hat \psi_j (k, \Delta t).
\label{Prod} 
\end{equation} 
We should stress that, by their definition as Fourier integrals, $\hat \psi_j (k, \Delta t)$ are given by complex functions of the wave vector $k$, and their appreciation as ``probabilities" has the only purpose of factorizing in Eq.~(\ref{Prod}). This factorized form is justified via the asymptotic matching procedure in the limit $k\rightarrow 0$. Without loosing in generality, it is sufficient to analyze a simplified version of Eq.~(\ref{Prod}) with only two processes included$-$one corresponding to a white noise-like process, which we shall mark by the index $L$; and the other one, corresponding to a regular convection process, such as a zonal flow or similar, which we shall mark by the index $R$. We have, accordingly,  
\begin{equation}
\hat \psi (k, \Delta t) = \hat \psi_{L}  (k, \Delta t) \hat \psi_{R} (k, \Delta t).
\label{Prod2} 
\end{equation}
These settings correspond to a set of Langevin equations
\begin{equation}
dx/dt = v;~dv/dt = -\eta v + F_R + F_L (t),
\label{Lvin} 
\end{equation}
where $\eta$ is the fluid viscosity; $F_R$ is the regular force; and $F_L (t)$ is the fluctuating (noise-like) force. We take $F_L (t)$ to be a white L\'evy noise with L\'evy index $\mu$ ($1 < \mu\leq 2$). By white L\'evy noise $F_L (t)$ we mean a stationary random process, such that the corresponding motion process, i.e., the time integral of the noise, $L (\Delta t) = \int_t^{t+\Delta t} F_L (t^{\prime}) dt^{\prime}$, is a symmetric $\mu$-stable L\'evy process with stationary independent increments and the characteristic function  
\begin{equation}
\hat \psi_L (k, \Delta t) = \exp (-K_\mu |k|^\mu \Delta t) \sim 1 - K_\mu |k|^\mu \Delta t.
\label{GCLT} 
\end{equation}
The last term gives an asymptotic inverse-power distribution of jump lengths 
\begin{equation}
\chi (\Delta x) \sim |\Delta x|^{-1-\mu}.
\label{Jump-l} 
\end{equation}
In the above, the constant $K_\mu$ constitutes the intensity of the noise. As is well-known, the characteristic function in Eq.~(\ref{GCLT}) generates L\'evy flights \cite{Klafter2004,Klafter}. 

Focusing on the regular component of the force field, $F_R$, it is convenient to represent the corresponding transition probability in the form of a plane wave, i.e.,
\begin{equation}
\hat \psi_R (k, \Delta t) = \exp (iuk\Delta t) \sim 1 + iuk\Delta t.
\label{Plane} 
\end{equation}
Here, $u$ is the speed of the ``wave," which is decided by convection. One evaluates this speed by neglecting the term $dv/dt$ in Langevin equations~(\ref{Lvin}) to give $u=F_R/\eta$. It is noted that the general condition in Eq.~(\ref{F2+}) is well satisfied for both the L\'evy processes and stationary convection, emphasizing the Markov property and the absence of trapping. Putting all the various pieces together, one obtains 
\begin{equation}
\hat \psi (k, \Delta t) = \exp (-K_\mu |k|^\mu \Delta t + ik F_R \Delta t / \eta).
\label{Tog} 
\end{equation} 
The next step is to substitute this into Eq.~(\ref{3}), and to allow $\Delta t \rightarrow 0$. Then, Taylor expanding on the left- and right-hand sides in powers of $\Delta t$, and keeping first non-vanishing orders, in the long-wavelength limit $k\rightarrow 0$ it is found that 
\begin{equation}
\frac{\partial}{\partial t} \hat f (k, t) = \left[-K_\mu |k|^\mu + ik F_R / \eta \right] \hat f (k, t).
\label{9} 
\end{equation} 
When inverted to configuration space, the latter equation becomes
\begin{equation}
\frac{\partial}{\partial t} f (x, t) =  \left[K_\mu \frac{\partial^\mu}{\partial |x|^\mu} - \frac{1}{\eta} \frac{\partial}{\partial x} F_R \right] f (x, t),
\label{Inv} 
\end{equation} 
where the symbol $\partial^{\mu} / \partial |x|^\mu$ is defined by its Fourier transform as  
\begin{equation}
\hat \mathcal{F} \Big\{\frac{\partial^\mu}{\partial |x|^\mu}f (x, t)\Big\} = -|k|^\mu \hat f (k,t).
\label{Def} 
\end{equation} 
In the foundations of fractional calculus (e.g., Ref. \cite{Samko}) it is shown that, for $1 <\mu < 2$, 
\begin{equation}
\frac{\partial^\mu}{\partial |x|^\mu} f (x, t) = \frac{1}{\Gamma_\mu}\frac{\partial^2}{\partial x^2} \int_{-\infty}^{+\infty}\frac{f (x^\prime, t)}{|x-x^\prime|^{\mu - 1}} dx^\prime.
\label{Def+} 
\end{equation} 
Equation~(\ref{Def+}) reproduces the Riesz fractional derivative discussed above, with $\Gamma_\mu = - 2\cos(\pi\mu/2)\Gamma(2-\mu)$. 

Relating $F_R$ to external potential field with the aid of $F_R = -V^{\prime}(x)$, and substituting in Eq.~(\ref{Inv}), one arrives at the following fractional Fokker-Planck equation, or FFPE (e.g., Refs. \cite{Ch2007,Klafter2004,Klafter,Gonchar}; references therein) 
\begin{equation}
\frac{\partial}{\partial t} f (x, t) =  \left[K_\mu \frac{\partial^\mu}{\partial |x|^\mu} + \frac{1}{\eta} \frac{\partial}{\partial x} V^{\prime}(x)\right] f (x, t).
\label{FFPE} 
\end{equation} 
Note that FFPE involves space fractional differentiation only in terms of the generalized Laplacian operator; whereas the convection term is {\it integer} and introduces the potential well for L\'evy flights. This observation elucidates the fundamentally different roles the stochastic and regular forces play as they set up the analytical structure of FFPE. In this context, the idea of ``fractional" convection term and some alternative generalizations of the Fokker-Planck equation (e.g., Ref. \cite{Report}) does not seem to find a solid dynamical background. FFPE in Eq.~(\ref{FFPE}) can alternatively be derived using as a starting point the set of Langevin equations~(\ref{Lvin}) instead of the evolution equation~(\ref{2}). The advantage of Langevin approach lies in the straightforward way of including the driving force terms in the presence of several competing dynamical processes in the medium. Previously, a study of nonlocal transport in terms of Langevin equations with L\'evy white noise and corresponding generalized Fokker-Planck equation containing space-fractional derivatives have been suggested by Fogedby \cite{Fog} and Jespersen {\it et al.} \cite{Jesper}. 

\subsection{The non-homogeneity issue} 
We should stress that the introduction of the $x$-dependent force $F_R (x) = -V^{\prime}(x)$ in place of the constant force in Eq.~(\ref{Inv}) destroys the spatial homogeneity of the transfer statistics implied by the transfer kernel in Eq.~(\ref{2}). Even so, this extension to non-homogeneous systems with the spatial asymmetry owed to the force $F_R = F_R (x)$ could be employed under the condition that the terms determining the jump length $|x-x^\prime|$ separate from the coordinate dependence in $F_R (x)$, implying that the force is calculated at the arrival site $x$ and not at the departure site $x^\prime$. Technically, the separation of terms can be implemented based on the generic functional form \cite{Barkai} of the memory kernel, using the Heaviside step function to ascribe the dependence on the jump length. More so, implementing a similar convention regarding the arrival site, the assumption that the intensity of the L\'evy noise $K_\mu$ does not depend on $x$ can be relaxed \cite{PLA}. In a basic physics perspective, the non-homogeneity is key to explain the occurrence of {\it superdiffusive} transport on combs and other subdiffusive structures, as the analysis of Ref. \cite{Baskin} has shown.

\subsection{Extension to Gaussian diffusion}
Equation~(\ref{FFPE}) can be extended, so that it includes {\it local} transport due to e.g., Coulomb collisions (as well as collisionless quasilinear transport), in addition to nonlocal transport processes discussed above. The key step is to observe that collisions$-$whatever nature they have$-$will generate a white noise process of the Brownian type, whose characteristic function is Gaussian and is obtained from the general L\'evy form~(\ref{GCLT}) in the limit $\mu\rightarrow 2$. Note that the Gaussian law, too, belongs to the class of stable distributions, but it will be the only one to produce finite moments at all orders. When the L\'evy and Brownian noises are included as independent elements to the dynamics, the transition probability in Eq.~(\ref{Prod}) will again factorize, and will acquire, in addition, a Gaussian factor $\hat \psi_G (k, \Delta t) = \exp (-D k^2 \Delta t)$, where $D$ has the sense of the diffusion coefficient. Then Eq.~(\ref{Tog}) will generalize to   
\begin{equation}
\hat \psi (k, \Delta t) = \exp (-K_\mu |k|^\mu \Delta t -D k^2 \Delta t + ik F_R \Delta t / \eta),
\label{CLT+} 
\end{equation}  
from which a FFPE incorporating both the Riesz fractional derivative and the usual Laplacian operator 
\begin{equation}
\frac{\partial}{\partial t} f (x, t) =  \left[K_\mu \frac{\partial^\mu}{\partial |x|^\mu} + D \frac{\partial^2}{\partial x^2} + \frac{1}{\eta} \frac{\partial}{\partial x} V^{\prime}(x) \right] f (x, t)
\label{FFPE+} 
\end{equation} 
can be deduced for $k\rightarrow 0$. Equation~(\ref{FFPE+}) reproduces the transport model in Eqs.~(\ref{MoDeL+}) and~(\ref{DL}) up to the sink terms in $\hat{S}_{\pm}[f (x, t)]$.  

\subsection{The boundary value problem}
The infinite limits of integration in the Riesz fractional derivative~(\ref{FDE}) and other fractional operators alike correspond to free L\'evy flights in open space. When placed on a comb, the L\'evy flyer will be subject to further restrictions owed to particularities of the potential force field, i.e., the shape of the $V(x)$ dependence. The focus here is on the jumps in $V(x)$ at each right border of the sawtooth (see Fig.~1). Those jumps would introduce infinite repulsive forces at $x=x_{j} +0$ for all $j=1,2, \dots$ starting from $x_j = 0$, making it impossible for the flyer to get back once it has crossed a tooth at some radial location $x=x_{j}$. The net result is that the transport process cannot propagate to the negative semi-axis because of the jump in $V(x)$ for $x\rightarrow +0$. If the spacing between the consecutive teeth of the comb is very broad, i.e., $\Lambda\rightarrow+\infty$, then we need to ensure there is no return at $x=0$. With this implication in mind, we limit the range of the integration in Eq.~(\ref{FDE}) to only half a space, i.e., $0 < x <+\infty$, advocating the following reduced form of the nonlocal function in Eq.~(\ref{DL}) (for $\mu\ne 1$) 
\begin{equation}
\Psi_\mu (x, t) = \frac{K_\mu}{\Gamma_\mu} \int_{0}^{+\infty}\frac{f (x^\prime, t)}{|x-x^\prime|^{\mu - 1}} dx^\prime. 
\label{Psi} 
\end{equation} 
Mathematically, this reduction of the limits of integration is important, as it provides consistency between the fractional integro-differentiation in FFPE and the sawtooth form of $V(x)$. Following Chechkin {\it et al.} \cite{Chechkin2003}, one finds in the presence of the no-return condition at $x=0$ that the transports model in Eqs.~(\ref{MoDeL+}) and~(\ref{DL}) with the $\Psi_\mu (x, t)$ function defined by Eq.~(\ref{Psi}) correctly phrases the first passage time density problem \cite{Ch2007,Klafter2004} for L\'evy flights. Moreover, this model will naturally observe the Sparre Andersen universality \cite{SA53} that the first passage time density decays as $\sim t^{-3/2}$ after $t$ time steps ($t\rightarrow +\infty$). We consider this universality as a characteristic property of the avalanche-diffusion transport system.  

\section{Analysis}
An important feature of Eq.~(\ref{MoDeL+}) is that it brings together processes occurring on kinetically disparate spatial scales ranging from the micro-scales of Coulomb collisions and/or electrostatic micro-turbulence to the meso-scales on which the shear flows organize themselves into a patterned staircase structure. It is understood that for $\Lambda\rightarrow+\infty$ the transport problem in Eq.~(\ref{MoDeL+}) is characterized by space scale separation in that there is a crossover scale, $\ell\ll \Lambda$, such that for $x \ll \ell$ the Gaussian diffusion (collisional and/or quasilinear-like) dominates; and for $x \gg \ell$ the nonlocal behavior dominates allowing for radially propagating avalanches and the Cauchy flights. The crossover scale $\ell$ is obtained by requiring that the Gaussian and the nonlocal terms in Eq.~(\ref{DL}) have the same order of magnitude, i.e., $D f(\ell, t) \sim \Psi_\mu (\ell, t)$ for $\mu\rightarrow 1$. This yields, with the aid of Eq.~(\ref{FDE+}), $\ell \sim \pi D/K_1$. Naturally, we require $\ell \ll \Lambda$ in the limit of strong turbulence. 

\subsection{Small scales: Collisional transport}
For $x \ll \ell$, we may neglect the second (nonlocal) term in Eq.~(\ref{DL}), keeping only the Gaussian term. Also for $x \ll \ell$ we may ignore the action of the potential field $V(x)$ in Eq.~(\ref{MoDeL+}), just remembering that it goes to zero for $x\rightarrow +0$ with its both first and second derivatives (owing to the condition $n>2$). Then from Eq.~(\ref{MoDeL+}) one sees that there is a steady-state ($\partial f (x, t) / \partial t = 0$; $f (x, t) = f(x)$) solution, which is determined by a bargain between the diffusion term, on the one hand, and the eventual sources and sinks, on the other hand, yielding,
\begin{equation}
- D \frac{\partial^2}{\partial x^2} f (x) = \hat{S}_{\pm}[f (x)]. 
\label{StSt} 
\end{equation} 
Next, we assume for simplicity, without loss of generality, that the sources $\hat{S}_{+}[f (x)]$ are delta-pulses centered at $x=x_j$. That means that $\hat{S}_{+}[f (x)] \equiv 0$ for $0 < x < \Lambda$. Concerning the sink terms, $\hat{S}_{-}[f (x)]$, we associate them with the stabilizing effect of the shear flows on radial transport \cite{Diamond} and the fact that such flows effectively absorb the particles (hence withdraw them from the radial diffusion processes) at a rate that is decided by the radial gradient of the intensity of the flow. In this regard, we may define $\hat{S}_{-}[f (x)] = -q f (x)$ for $x \ll \ell$, where $q$ is a coefficient, which characterizes the efficiency of the absorption. Then from Eq.~(\ref{StSt}) one finds that the decay of the probability density is exponential, i.e., $f (x) \sim \exp (- \sqrt{q/D}\, x)$, with a characteristic decay length of the order of $\sqrt{D/q}$. Consistently with the above reasoning, we require $\sqrt{D/q} \lesssim \ell \ll \Lambda$. 

\subsection{Long scales: Nonlocal transport}
The dynamical picture changes, if the spatial scale $x$ overshoots $\ell$, i.e., $x\gg\ell$. In this parameter range, the diffusion term may be neglected, as it will be much smaller than the competing L\'evy term. Also, because the avalanches propagate radially on a very fast time scale, if not at all ``instantaneously," their continuum damping by the shear flows in-between the staircase spikes will be rather unimportant (at contrast to local diffusion), making it possible to omit the sink term in Eq.~(\ref{MoDeL+}). Then the auspicious steady-state solution is defined through a negotiation between the nonlocality contained in the L\'evy term, on the one hand, and the fluid nonlinearities generating the potential function $V(x)$, on the other hand. With the aid of Eqs.~(\ref{FDE}) and~(\ref{Psi}), one gets
\begin{equation}
- \frac{1}{\eta} \frac{\partial}{\partial x} V^{\prime}(x) f (x) = \frac{K_\mu}{\Gamma_\mu} \frac{\partial^2}{\partial x^2} \int_{0}^{+\infty}\frac{f (x^\prime)}{|x-x^\prime|^{\mu - 1}} dx^\prime.
\label{VNL} 
\end{equation}      
Using here that the total probability is conserved across the integration domain, i.e., $\int_0^{+\infty}f (x^\prime) dx^\prime = 1$, one infers the following asymptotic matching condition for the function $f(x)$ in the limit $x\rightarrow+\infty$, that is, $V^{\prime}(x) f(x) \propto x^{-\mu}$. 
Recalling further that the leading term in the expansion of $V(x)$ goes as a power-law, i.e., $V(x) \propto x^n$, with $n>2$, one gets for $x\rightarrow+\infty$ 
\begin{equation}
f(x) \sim (\eta K_\mu / \Gamma_\mu)\, x^{-(n+\mu - 1)}.
\label{Tail} 
\end{equation}    
Note that there is no algebraic tail for $\mu\rightarrow 2$ because of the divergence $\Gamma_\mu\rightarrow+\infty$. For $\mu < 2$, we require that the probability density $f (x)$ decays faster than {\it any} L\'evy stable law, that is, faster than the inverse-cube dependence $\propto x^{-3}$ in the limit $x\rightarrow+\infty$ \cite{Ch2007,Klafter}. That would mean that the second moments become {finite} in the presence of the potential field $V(x)$, i.e., $\int_0^{+\infty} {x^\prime}^2 f (x^\prime) dx^\prime < +\infty$. Then the finiteness of the second moments would imply in turn that the avalanches are asymptotically localized in the sense of L\'evy-Gnedenko generalized central limit theorem \cite{Gnedenko}. So, the localization condition is, essentially, a condition on the $n$ value and reads 
\begin{equation}
n+\mu - 1 > 3,
\label{Cond} 
\end{equation}    
that is, $n > 4-\mu$. In the case of Cauchy flights, we have $n > 3$ (in view of $\mu\rightarrow 1$). The net result is that the Cauchy flights are asymptotically localized by a potential field $V(x)$, whose leading power grows faster than $\propto x^3$ for $x\rightarrow+\infty$. If $n$ is integer, then the condition $n > 3$ implies it is the bi-quadratic dependence $\propto x^4$ that localizes the Cauchy flights in the lowest order.

\section{Discussion}
Our findings so far can be summarized as follows. The avalanche-diffusion model in Eqs.~(\ref{MoDeL+}) and~(\ref{DL}) is characterized by space scale separation, so that at the short scales (shorter than the crossover distance $\ell \sim \pi D/K_1$) the transport is dominated by ordinary (Brownian-like) diffusion processes, and at the far longer spatial scales it is dominated by nonlocal phenomena involving plasma avalanches. The latter are coherent structures mediating the Cauchy flights of passive particles in radial direction, with the fractal dimension $\mu\rightarrow 1$. The decay of the probability density in a steady state of the coupled avalanche-zonal flow system is exponential within the diffusion domain and is inverse power-law in the nonlocal domain. The exponent of the power-law is $-(n + \mu - 1)$ and is defined by the leading term in the $V(x)$ expansion for $x\rightarrow+\infty$. In the above we have requested that $n$ be larger than 2, which was motivated by the mathematical structure of the convection term on the left of Eq.~(\ref{MoDeL+}). For $2 < n \leq 3$, the decay of the probability density corresponds to a L\'evy stable law, with diverging second moments, and the avalanches appear to be {\it not} localized. On the contrary, for $n >3$, the probability density vanishes faster than the steepest L\'evy stable law would decay. This reinstalls finiteness of the second moments implying that the avalanches are asymptotically localized. Thus, there is a critical dependence in the $V(x)$ function, i.e., the cubic dependence $\propto x^3$, such that for dependences faster than this the nonlocal features are confined at the staircase steps, and will be unconfined otherwise.  

\subsection{Finite-size effects}
In the above we have assumed that the $\Lambda$ value is actually very large, and we have neglected accordingly any finite size effects$-$to be attributed to the fact that the probability density $f (x)$ might not have completely vanished yet before the next tooth of the comb is faced. To this end, because of the sharp drop-off in the $V(x)$ dependence at $x=\Lambda +0$, there may be an important probability of barrier crossing, so that the avalanches having finite inertia would just tunnel under the barrier. If the barrier is successfully crossed, then in the idealized model the dynamics is reset to the next step of the comb, with an updated boundary condition, and the process repeats itself. One sees that there will be net transport in radial direction propagating to long distances, and this occurs in ordered steps along the $x$ axis, with the characteristic step $\Lambda$. The process can be thought as a persistent random walk down to the scrape-off layer, with a bias posed by the asymmetry of the comb's teeth. Theoretically, it corresponds to the transport case with finite moments and superdiffusive scaling and has been considered for combs in Ref. \cite{Baskin}. It is understood that in the presence of a characteristic step-size there is no asymptotic nonlocal behavior in the L\'evy-Gnedenko sense, even though the entire process is not confined in the long run. These complex features have been seen in simulations \cite{DP2017,Hornung,Wang}. 

Our next point here concerns the absence of power-law tails in the Gaussian limit $\mu\rightarrow 2$, as Eq.~(\ref{Tail}) has shown. In this case, the decay of the $f(x)$ function is exponential through the entire staircase period, i.e., $f (x) \sim \exp (- \sqrt{q/D}\, x)$ for all $0 < x\lesssim\Lambda$. Since $\Lambda\gg\sqrt{D/q}$, the exponential factor is quite small at $x\rightarrow\Lambda -0$. Hence, the probability of barrier crossing is negligible, implying that (i) the transport process is well localized within the barrier, and (ii) there is no net transport at the macroscopic scales (beyond the staircase period). This conclusion substantiates the result of del-Castillo-Negrete {\it et al.} \cite{DCN}, who associated the absence of transport at a macroscopic level with a dynamical system reaching local thermodynamic equilibrium for $\mu \rightarrow 2$.  

Self-consistently, one would expect that the staircase patterning and the generation of L\'evy noises in the medium are two faces of the same coin, that is, two coupled processes operating in the same complex system far from thermodynamic equilibrium. If this conjecture is correct, then (i) triggering transport barriers in magnetically confined plasma unavoidably generates transport avalanches contesting these barriers; (ii) transport models not including nonlocal phenomena in the medium are inadequate to describe the staircase self-organization; (iii) $\mu$ may be taken as a measure of how far from equilibrium the dynamical system is \cite{UFN}; and (iv) transport in preferred direction has parametric dependence on $\mu$ and is intensified, if the $\mu$ value is lowered. This parametric behavior has been confirmed numerically \cite{DCN}.

\subsection{Size distribution of avalanches}
In large systems, the avalanches being coherent structures may have a nontrivial size distribution, and this may be obtained as the probability for the random walker to {\it not} be dispersed by the Fokker-Planck dynamics after $\Delta s$ space steps in radial direction, enabling
\begin{equation}
w(\Delta s) = \left[\int_{0}^{+\infty} - \int_{0}^{\Delta s}\right] f (x^\prime)dx^\prime = \int_{\Delta s}^{+\infty} f (x^\prime)dx^\prime.
\label{Size} 
\end{equation}    
Note that the conservation law $\int_0^{+\infty}f (x^\prime) dx^\prime = 1$ implies $\lim_{\Delta s\rightarrow 0} w(\Delta s) = 1$. Utilizing the corresponding representations for the $f(x)$ dependence in both core (small scales: Sec. III-A) and tail (long scales: Sec. III-B) regions, and integrating in Eq.~(\ref{Size}) from $\Delta s$ to $+\infty$, one finds that the size distribution of avalanches $w(\Delta s)$ interpolates between the initial exponential form $w(\Delta s) \sim \exp (- \sqrt{q/D}\, \Delta s)$ for $\Delta s \ll \ell$ and the asymptotic inverse power-law behavior 
\begin{equation}
w(\Delta s) \sim (1 / \Gamma_\mu)\, \Delta s^{-(n+\mu - 2)}
\label{Tail+} 
\end{equation}    
for $\Delta s \gg \ell$. In the above we have promoted the gamma function to emphasize that there is no asymptotic power-law behavior in the Gaussian limit, $\mu\rightarrow 2$. In case of the bi-quadratic ($n=4$) dependence in the leading order, one gets, using Eq.~(\ref{Tail+}), $w(\Delta s) \sim (1/\Gamma_\mu)\, \Delta s^{-(2+\mu)}$ for $\Delta s \gg \ell$. In particular, for the Cauchy flights, with $\mu = 1$, $w(\Delta s) \propto \Delta s^{-3}$. 

\subsection{Connection to the Weibull distribution and other distributions with long tails}
On the experimental/modeling side, the size distribution $w(\Delta s)$ has already been measured in computer simulations of the Tore Supra plasma \cite{DP2017}. The results deriving from those measurements have been plotted against the Fr\'echet distribution, which is a special case of the Weibull (or generalized extreme value) distribution with lower bound. A summary of this analysis is given by Eq.~(3) of Ref. \cite{DP2017}, yielding the analogue $w(\Delta s)$ function deduced phenomenologically from the simulations. By examining the result of Ref. \cite{DP2017} one sees that the Weibull distribution reproduces both the exponential (small sizes) and the power-law (large sizes) counterparts of the $w(\Delta s)$ dependence and in this sense offers qualitative agreement with the limiting cases of the avalanche-diffusion model discussed above. 

Quantitative agreement is obtained by matching the exponent of the algebraic tail of the Weibull distribution (in the notation of Ref. \cite{DP2017}, this exponent is written as $-(1+\kappa)/\kappa$, where $\kappa$ is numerical fitting parameter) to our $-(n+\mu - 2)$ in Eq.~(\ref{Tail+}). The result is the matching condition $n = (3-\mu) + 1/\kappa$. Using $\kappa \approx 0.6$ as of Ref. \cite{DP2017}, and setting the index $\mu$ to unity, one obtains $n \approx 3.7$. So, the effective value of $n$ fitting the data is clearly greater than 3, with a fair margin. This implies localization, and this in fact has been observed \cite{Hornung,DP2017}. 

The Weibull distribution discussed in Ref. \cite{DP2017} is analytically very similar to the so-called ``kappa" distribution, which has come of age as a suitable phenomenological fitting tool when describing dynamic phenomena in complex systems (e.g., Refs. \cite{UFN,Chapter,Sharma}; references therein). The theoretical significance of the kappa distributions lies in the fact \cite{NPG} that these distributions appear as canonical distributions in the non-extensive thermodynamics due to Tsallis \cite{Tsallis88}. There have been some discussion in the literature concerning the possible relationship between the Tsallis entropy and L\`evy flights (e.g., Ref. \cite{Alemany}). Here, we might partially support that discussion, however, we draw attention to the fact that the L\`evy flights alone are not sufficient to generate the kappa distributions, and one needs, in addition, a process producing the exponential decay part at the microscopic scales. This is accounted for by the sink term in $\hat{S}_{-}[f (x)]$, which is motivated in our model by the stabilizing effect of the shear flows on radial diffusion, and which has been written as $\hat{S}_{-}[f (x)] = -q f (x)$ for $x\ll\ell$.  

\subsection{Black swans}
When a passive particle is caught on an avalanche, it gains a kick of kinetic energy, and we have tacitly assumed that this energy being possibly large in absolute terms is, however, small compared to $V_{\max} = V(\Lambda)$. This assumption was guaranteed by $\Lambda\rightarrow+\infty$ permitting a steady state solution for the probability density $f (x)$, with the Sparre Andersen universality \cite{SA53} dictating the reduced limits of integration in Eq.~(\ref{VNL}). Then it was our conclusion that the avalanches could be effectively confined within the staircase steps, provided just that the power $n$ in the shape function $V(\Delta x) \propto |\Delta x|^n$ is greater than 3 (see Fig.~1). 

In a magnetically confined plasma, the coupled avalanche-zonal flow interacting system may behave similarly to a predator-prey system in that the transport barriers generated by the turbulence take energy from the turbulence, meaning that their driving mechanism is diminished, and they may be decaying due to classical or neo-classical collisional damping (e.g., Refs. \cite{Diamond,Schmitz}). The process opens a possibility that some avalanches escape the confinement domain during the barrier depression periods, giving rise to sporadic bursts of large-scale transport well above the staircase's parapet. This type of occasionally strong transport events being virtually insensitive to the underlying flow and stress organization has been found in the GYSELA simulations \cite{DP2017,Hornung}, and their statistical weight has been assessed to be about a percentile of all avalanche events observed across the staircase. 

If one is a traditionalist, and wants to remain with the Fokker-Planck model in Eqs.~(\ref{MoDeL+}) and~(\ref{DL}), then one might readily assess the statistical case of unconfined avalanches as follows. In the basic kinetic equations, one neglects both the Gaussian and the potential force terms, as well as the sink term $\hat{S}_{-}[f (x)]$, and only keeps the nonstationary term against the L\'evy term. The net result is that (i) there is no steady state solution, contrary to the confined transport case; and (ii) the probability density, which is time dependent, behaves asymptotically as a power-law $f(x, t)\sim K_\mu t / x^{1+\mu}$. Due to this property, the mean squared displacement diverges, i.e., $\langle x^2 (t)\rangle\rightarrow+\infty$, which is typical for free L\'evy flights. In view of this divergence, the size distribution of unconfined avalanches is obtained as the corresponding jump length distribution \cite{Klafter}. The latter is given by Eq.~(\ref{Jump-l}), yielding, for $\Delta s \gg \ell$, $\Delta s \gg \sqrt{D/q}$,
\begin{equation}
w(\Delta s) \propto \Delta s^{-(1+\mu)}.
\label{Black} 
\end{equation} 
The scaling in Eq.~(\ref{Black}) is confirmed by tuning $n$ to its borderline value $n=3$ in $w (\Delta s) \propto \Delta s^{-(n+\mu - 2)}$, as is intimated by Eq.~(\ref{Cond}) above. 

Let us christen our avalanches. Inspired by the mathematical elegance of the confined L\'evy flight, we baptize the avalanches caught in-between the staircase steps {\it white swans}. The term is intended to contrast the other population of bursty transport events, the {\it black} swans, which are the avalanches escaping the confinement system during the low barrier phase. The name {\it black swan} is borrowed from the Taleb's book \cite{Taleb}; where, it has been introduced to describe an unexpected catastrophic event catching us off-guard. Note that the size distributions of the power-law type appear for both the white and black swans, but with different drop-off exponents, so that for $n > 3$ the black-swan distribution is {\it always} flatter (in its habitat) than the corresponding white-swan distribution (see Fig.~2). 

The occurrence of the black-swan family gives rise to a characteristic ``bump" in the $w(\Delta s)$ dependence, which is located around $\Delta s \sim\Lambda$. Given the space scale separation condition $\Lambda\gg \sqrt{D/q}$, the position of this bump is well beyond the exponential core region (see Fig.~2). One sees that the resulting $w(\Delta s)$ dependence, which embraces both the white- and black-swan populations, will be {\it bi-modal} in that it has a second maximum near $\Delta s \sim\Lambda$. 

Note, also, that the white swans go extinct beyond the staircase spacing distance $\sim\Lambda$, that is, the areas of the white- and black-swan dominance are essentially different (except for the narrow overlap region around $\sim\Lambda$). This finding is peculiar and says the probabilities of the black-swan events {\it cannot} be predicted by interpolating the white-swan counterpart (if it exists) to longer sizes. 

The respective drop-off exponents for the white and black swans would only coincide for the borderline case $n=3$, for which all the swans stick together to form one single family, with the unique size distribution $w(\Delta s) \propto \Delta s^{-(1+\mu)}$. Arguably, one might refer to this borderline case as {\it grey} swans, as they serve as the missing bond between the two main species, the white and black swans. Because $\mu < 2$, the grey swans correspond to non-localized avalanches. 

For $n < 3$ (but still larger than 2, see Sec. II), we expect the white swans to completely change their color and ``mutate" (past the intermediate grey-swan phase) into one single family of the black-swan type populating the entire staircase (see Fig.~3), with the unique size distribution $w(\Delta s) \propto \Delta s^{-(n+\mu -2)}$. As this ``mutation" occurs, the bump around $\Delta s \sim\Lambda$ disappears. We associate this with the loss of bi-modality and related bifurcation phenomena studied by Chechkin {\it et al.} \cite{Chechkin2002}. This regime shift could be interpreted as a localization-delocalization transition \cite{EPL_2012} on the comb structure shown in Fig.~1. 

\begin{figure}
\includegraphics[width=0.55\textwidth]{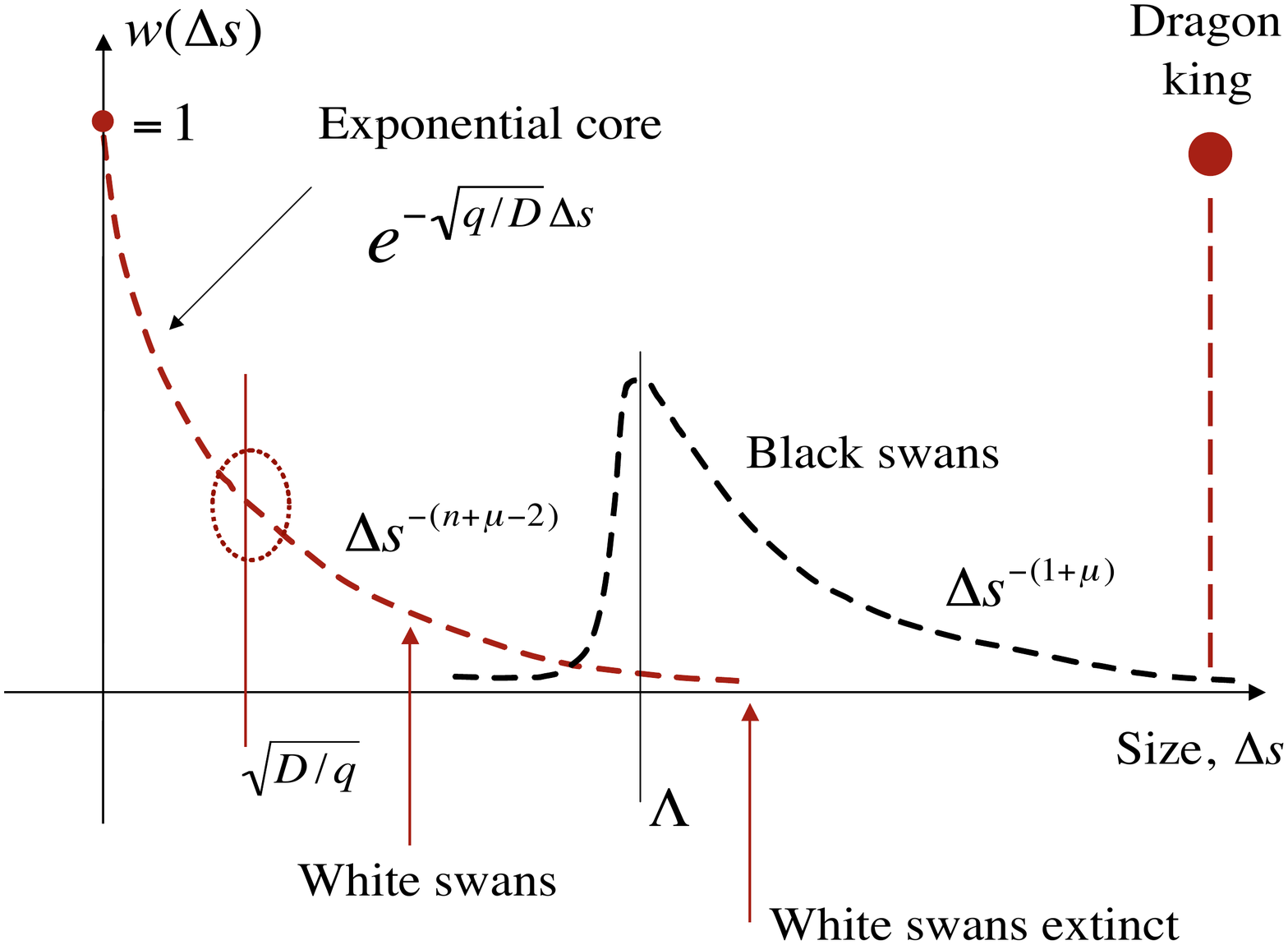}
\caption{\label{} The coexistence between the white- and black-swan families of avalanches for $n > 3$. The occurrence of the black-swan family gives rise to a characteristic ``bump" in the $w(\Delta s)$ dependence around $\Delta s \sim \Lambda$, lying far off the exponential core region (i.e., the property of bi-modelity). The dragon-king avalanches being singular transport events are shown as a fat dot at the upper-right corner dominating the scene.}
\end{figure}
\begin{figure}
\includegraphics[width=0.55\textwidth]{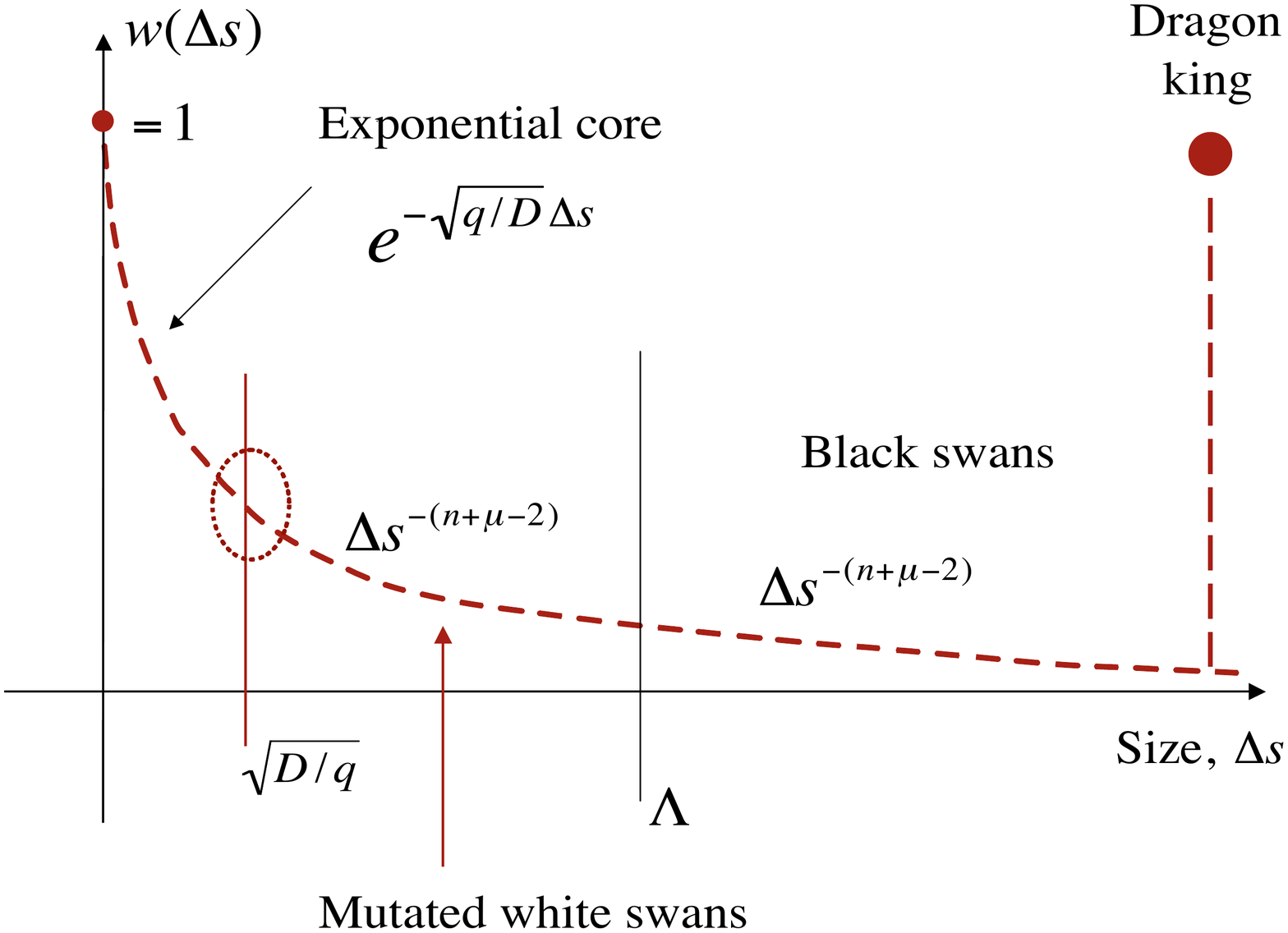}
\caption{\label{} Same situation, but for $2 < n < 3$. The regime with $n=3$ is the borderline case, for which the white-swan family ``mutates" into one extended black-swan family past the grey-swan species. The bi-modelity of the $w(\Delta s)$ dependence (see Fig.~2 above) is naturally lost in this case.}
\end{figure}

If one starts from poor confinement, with the black swans being the dominant species, and intervenes on the $n$ value trying to bring it above the $n=3$ border, then one encounters a bifurcation point, at which one witnesses the occurrence of a new family of avalanches, the white swans, which is the ``mutation" of the black swans trapped in-between the staircase steps. Past the bifurcation point at $n=3$, the $w(\Delta s)$ function becomes bi-modal, with a distinct, steeply decaying branch in the subrange $\Delta s \lesssim \Lambda$ (the white swans), and the asymptotic black-swan behavior for $\Delta s \gg \Lambda$, with a drop-off exponent conforming with a L\'evy stable law. If $n$ is integer, then the white swans would be identifiable starting from $n\geq 4$, i.e., when the growth of $V(\Delta x)$ is bi-quadratic ($n=4$) in the leading order. Tuning the fractal dimension $\mu$ to 1, we have for $n=4$, $w(\Delta s)\propto \Delta s^{-3}$ in the white-swan category, and $w(\Delta s)\propto \Delta s^{-2}$ in the black-swan category. The two populations are quite separate in this case (see Fig.~2) and, moreover, fairly divide their habitats in that the white swans reign in the domain $\sqrt{D/q} \ll \Delta s \lesssim \Lambda$ and the black swans reign in the domain $\Delta s \gg\Lambda$. The fact that the black swans had adhered to an $\propto \Delta s^{-2}$ drop-off might be substantiated by the analysis of Ref. \cite{Sornette_PRL}, in which the dynamics of coupled chaotic oscillators with extreme events was investigated numerically.   

We should stress that the black swans occupy the most ``dangerous" niche corresponding to large-amplitude events, with sizes generally greater than $\sim\Lambda$. In a practical advisory, that may mean the following. The statistics of large-amplitude bursts of transport (the black swans) may be quite different from the statistics of smaller events (as much as the difference between black and white). So, if one wants to predict the transport at the macroscopic (system-size) scales, then one {\it cannot} really interpolate from meso-scales to the large scales along the white-swan branch, as that would miss the important population of the black swans coming up. Indeed, ``More is different" \cite{And} for complex systems, and this is illustrated even further in Sec. IV-E. 

\subsection{Dragon kings}
The swans whatever color they have won't be the unique species of the avalanche events populating the staircase. In strong drift-wave turbulence, there is an important probability that the avalanches themselves are sources of secondary instabilities, and these would merge with the mother instability via inverse cascade of spectral energy, giving rise to ever amplifying unstable fronts propagating radially toward the scrape-off layer \cite{PLA,JPP}. The amplification occurs when the Rhines time in the system is small compared with the instability growth time. We note in passing that the Rhines time \cite{Naulin} in drift-wave turbulence is the ratio between the Rhines length (which is proportional to the square-root of the $E\times B$ velocity) and the $E\times B$ velocity itself, i.e., the decay of the Rhines time is given by the {\it inverse} square-root of the $E\times B$ drift. Clearly, the smallness of the Rhines time implies that the fluctuations are strong, and the turbulence level high. One sees that the avalanche is amplified, because it induces {\it secondary} turbulence on its front and simultaneously absorbs this turbulence through the inverse cascade enhancing the instability. The result of this amplification (and amplification of the amplification, etc.) is an avalanche of extraordinarily great size, washing out all the finer scale structures on its way down to the scrape-off layer. These stark events would be ``true" extreme events in our system, and their energy content is only limited to the system size. There have been a mythic term to define such events for complex systems, {\it dragon kings}, which have been introduced by Sornette \cite{Sornette} to emphasize their superiority over any other transport event around. A defining feature of the dragon-kings (other than their ``noble" rank) is the fact that they do not belong to the typical power-law branch representing the black swans, but would, rather, keep away from the mainstream statistics, being a restricted family of ``odd" events of anomalously large magnitude (and the associated rare appearance).  

In a statistical perspective, the interest in dragon kings lies in the fact that they represent extreme events beyond the usual scale-free paradigm, and their occurrence frequencies are much higher than what would be expected under a power-law approximation to  the correspondingly great sizes. When drawn to the probability density-size diagram, the dragon kings would appear as a peak at the right corner of the black-swan distribution (see the schematic illustrations in Figs.~2 and~3), such that the probability mass under the peak corresponds approximately to the integral of the probability density that would result if the black-swan population extended to infinity \cite{Sornette_PRL}. A summary on current scientific debate concerning the issue of dragon kings, and the methods to detect them, can be found in a Topical review in Ref. \cite{Sornette_2012}. Direct experimental evidence of large amplitude avalanche events at the edge of the JET plasma has been reported by Xu {\it et al.} \cite{Xu}. 

Given for granted that the dragon-king avalanches have outstanding expect size, we disregard the idea these avalanches may be described under the Fokker-Planck dynamics in Eqs.~(\ref{MoDeL+}) and~(\ref{DL}). Theoretically, this makes the situation unavoidably more debatable and controversial. As a prospective model approach, one might tackle a complex system with mixed multiscale-coherent behavior \cite{Chapter,Sharma}. In such systems, one often finds that there is a subordination between the different order parameters, that is, the multi-scale ordering generating the power-law branch (black swans) may act as input control parameter for the emerging coherent ordering \cite{Chapter}. This {\it competition} between the two orderings may result in an explosive instability in the system (i.e., the ``blow-up" of phase space trajectories generating a dragon king-like event) and mathematically corresponds to a description in terms of fractional Ginzburg-Landau equation \cite{UFN,FGL}. An alternative approach discussed in Ref. \cite{PPCF06} has used the idea of complex nonlinear Scr\"odinger equation with integer derivatives, in which the free energy source term was coupled to the nonlinear term, giving rise to the phenomena of convective amplification and ballistic radial propagation of unstable fronts (our dragon-king avalanches). 

The ``blow-up" of phase-space trajectories in a system of coupled chaotic oscillators with master-slave subordination and transverse instability has been demonstrated numerically in Ref. \cite{Sornette_PRL}. In these simulations, the blow-up occurred when the trajectories occasionally touched on ``hot spots" of the chaotic system with a highly inhomogeneous phase space. It has been discussed that the blow-up$-$also termed {\it attractor bubbling}$-$could be directly responsible for the occurrence of dragon kings in this specific configuration, and that the dragon kings, in general, are likely in networks of coupled nonlinear oscillators with subordination \cite{Sornette_PRL,Cross-D}. 

A model of explosive instability considered by Eliazar in Ref. \cite{Eliazar} suggests the dragon kings and black swans may appear universally and jointly through dynamics. He argued the black-swan branch could be an indication that the dragon kings are but exploded black swans and may materialize even in deterministic systems under special initial conditions. 

All in all, these observations may have important implications for the dynamics of coupled drift wave-zonal flow-avalanche system, for which one might expect outstanding bursts of transport beyond the black-swan metrics \cite{Striking}.

\section{Concluding remarks}
{In summary}, we have shown that a potential function that grows steeply enough with the spatial scale may confine nonlocal transport with L\'evy flights. This finding has important implications for the understanding of localization-delocalization phenomena in banded flows observed in planetary atmospheres \cite{McIn,Markus}, terrestrial oceans \cite{Ocean}, and, more recently, in tokamak plasma \cite{DP2015,Hornung,DP2017,Wang}. Also it offers a simple criterion to characterize internal transport barriers that may or may not confine the nonlocal transport. We have discussed that the nonlocal features could be introduced by so-called transport avalanches, which may trap and convect particles in radial direction at about a sonic speed. A mixed avalanche-diffusion model for the probability density produces the size distribution of avalanches in qualitative (and given the fitting parameter kappa, also quantitative) agreement with observations. 

Further focusing on the phenomena of localization-delocalization (and the associated power-law reduced drop-off of the probability density), we have discussed that there may exist different families of avalanches populating the plasma staircase, and we have theoretically predicted at least three such families depending on the dynamical features they represent: (i) the {\it white swans}, i.e., the avalanches confined in-between the staircase steps; (ii) the {\it black swans}, i.e., the avalanches that may occasionally escape the confinement domain as a result of the predator-prey dynamics of the coupled avalanche-zonal flow system (or other nonlinear phenomena alike); and (iii) {\it dragon-kings}, i.e., events of extraordinarily large magnitude, which represent the catastrophic events in the system, with possible irreversible consequences. We expect the black swans to be the dominant population particularly during the phases of barrier lowing posed by the predator-prey oscillation of the turbulence patterns in magnetic confinement geometry \cite{Diamond,Schmitz}. At contrast, dragon kings likely afford a different evolution path related with the phenomena of induced vortex formation \cite{PLA,JPP} and amplification (and amplification of the amplification, etc.) of secondary instabilities in the presence of inverse spectral energy cascade. Concerning the white-swan population, it {only} appears in the model, if the potential function $V(\Delta x)$ grows faster than $\propto |\Delta x|^3$ in the leading order, and is totally absorbed by the expanding black-swan family otherwise. This gives rise to a localization-delocalization transition at the cubic dependence $V(\Delta x) \propto |\Delta x|^3$ and the  associated loss of bi-modality consistently with the results of Refs. \cite{Chechkin2002,Klafter2004}. If one is precise and happens at the transition point exactly ($n=3$), then one finds (iv) the elusive {\it grey swans}, which represent the connecting bond between the white- and black-swan species, and which are {\it not} localized, with the size distribution $\propto \Delta s^{-(1+\mu)}$ conforming to a L\'evy stable law. We have proposed that both the white and black swans could be described in terms of the Fokker-Planck model with a comb-like potential force term and properly defined nonlocal term; whereas the dragon kings being exceptionally strong events of explosive type corresponded to a different description advancing the notion of fractional Ginzburg-Landau equation \cite{Chapter,FGL,UFN}. The results, presented in this work, pave the way for the construction of a {\it self-consistent} theory of nonlocal transport, according to which the avalanches are localized (or not localized) by the same comb-like potential field that generates these avalanches. This proposal might breath new life into the work in Ref. \cite{Baskin}, in which the occurrence of L\'evy-like processes on subdiffusive structures has been considered. Further research in this direction might be strongly advocated.       

\acknowledgments
The authors thank G. Dif-Pradalier and the participants of the 9$^{\rm th}$ Festival de Th\'eorie in Aix-en-Provence for many interesting discussions. One of the authors (A.V.M.) acknowledges the hospitality and partial support at the International Space Science Institute (ISSI) at Bern, Switzerland, during June, 2018, as well as constructive and stimulating discussions with R. Rodrigo and the ISSI visitors. Also A.V.M. thanks D. Sornette for sharing insights into the topic of dragon king, and for highlighting the works in Refs. \cite{Sornette_PRL,Cross-D}. This study has been carried out within the framework of the EUROfusion Consortium and has received funding from the Euratom research and training programme 2014-2018 under grant agreement No 633053 for the project AWP17-ENR-ENEA-10. The views and opinions expressed herein do not necessarily reflect those of the European Commission.

%
%
%
%


\begin{thebibliography}{}
\bibitem{Striking}
A challenging example of an outlying event of anomalously large size resembling a dragon-king has been developed recently by G. Dif-Pradalier based on flux-driven gyrokinetics, using the GYSELA code with realistic Tore Supra plasma parameters.
\end{thebibliography}

\end{document}